\documentclass{iaus}
\usepackage{graphicx}

\title[Preliminary Results from the NASA {\em EPOXI} Mission] %% give here short title %%
{Preliminary Results on HAT-P-4, TrES-3, XO-2, and GJ 436 from the NASA {\em EPOXI} Mission}

\author[Ballard et al.]   %% give here short author list %%
{Sarah Ballard$^1$,
 David Charbonneau$^1$,
 Michael F.~A'Hearn$^2$,
 Drake Deming$^3$,
 Matthew J.~Holman$^1$,
 Jessie L. Christiansen$^1$,
 David~T.~F.~Weldrake$^1$,
 Richard K.~Barry$^3$,
 Marc J.~Kuchner$^3$,
 Timothy~A.~Livengood$^3$,
 Jeffrey Pedelty$^3$,
 Alfred Schultz$^3$,
 Tilak Hewagama$^2$,
 Jessica M.~Sunshine$^2$,
 Dennis D.~Wellnitz$^2$,
 Don L.~Hampton$^4$,
 Carey~M.~Lisse$^5$,
 Sara Seager$^6$,
 Joseph F.~Veverka$^7$}

\affiliation{$^1$Harvard-Smithsonian Center for Astrophysics, \\ 60 Garden Street, Cambridge, MA 02138, USA\\ email: {\tt sballard@cfa.harvard.edu} \\[\affilskip]
$^2$University of Maryland, College Park, MD 20742, USA\\[\affilskip]
$^3$Goddard Space Flight Center, Greenbelt, MD 20771, USA\\[\affilskip]
$^4$University of Alaska Fairbanks, Fairbanks, AK 99775, USA\\[\affilskip]
$^5$Johns Hopkins University Applied Physics Laboratory, Laurel, MD 20723, USA\\[\affilskip]
$^6$Massachusetts Institute of Technology, Cambridge, MA 02159, USA\\[\affilskip]
$^7$Cornell University, Space Sciences Dept, Ithaca, NY 14853, USA\\}

\pubyear{2008}
\volume{253}  %% insert here IAU Symposium No.
\pagerange{1--4}
\jname{Transiting planets}
\editors{A.C. Editor, B.D. Editor \& C.E. Editor, eds.}
\begin{document}

\maketitle

\begin{abstract}
EPOXI (EPOCh $+$ DIXI) is a NASA Discovery Program Mission of Opportunity using the Deep Impact flyby spacecraft. The EPOCh (Extrasolar Planet Observation and Characterization) Science Investigation will gather photometric time series of known transiting exoplanet systems from January through August 2008. Here we describe the steps in the photometric extraction of the time series and present preliminary results of the first four EPOCh targets. 
\keywords{techniques: photometric, planetary systems, planets and satellites: general}
%% add here a maximum of 10 keywords, to be taken form the file <Keywords.txt>
\end{abstract}

\section{Introduction}
The high-precision time series produced from the EPOCh observations, with their substantial time and phase coverage, allow us to pursue three important science goals with respect to transiting exoplanets. First, we will search for evidence of reflected light from the giant transiting planets at secondary eclipse (Rowe et al. and Matthews et al., these proceedings). Second, EPOCh will be sensitive to terrestrial planets in these systems, using two detection methods. The first will be a search for direct transit events caused by approximately Earth-sized planets, particularly ``hot Earths'' that are predicted to be captured in low order mean motion resonances with the giant planets (\cite[Croll et al. 2007a,b]{Croll2007a}). EPOCh will also perform transit timing measurements for the giant planets, and search for timing variations due to terrestrial planet perturbers (\cite[Agol et al. 2005]{Agol05}; \cite[Holman \& Murray 2005]{Holman05}). The third science goal of the EPOCh mission is a precise determination of the system parameters, notably the planet radius, and a search for planetary satellites and circumplanetary rings through their perturbations to the transit light curve (\cite[Brown et al., 2001]{Brown01}; \cite[Barnes \& Fortney, 2004]{Barnes04}). These proceedings will summarize the steps we take to produce the time series, as well as describe the preliminary search for additional transiting bodies. 
 
Each time series consists of continuous 50-second integrations through a clear visible filter with the 30 cm aperture High Resolution Imager (HRI) instrument. These observations have a typical duration of 3 weeks per target. The spacecraft will then continue en route to its 2010 encounter with comet Hartley-2, the target of the DIXI Science Investigation. For a history of the EPOCh observations and more information about the EPOXI mission, please see Christiansen et al. (these proceedings).
%\end{section}

\section{Data Reduction}
We use the existing Deep Impact reduction pipeline to perform bias and dark subtraction, as well as flat fielding (\cite[Klaasen et al. 2005]{Klaasen05}). We then conduct the photometric extraction as follows. We determine the position of the star on the CCD using PSF fitting, by maximizing the goodness-of-fit (with the $\chi^{2}$ statistic as an estimator) between an image and a model PSF with variable position, additive sky background, and multiplicative brightness scale factor. The PSF itself, with a roughly 10 pixel FWHM, is produced from the drizzle of more than 1200 60$\times$60 pixel cutouts, filtered to eliminate cosmic ray hits before drizzle. The final PSF product is sampled to a tenth of a pixel. A bilinear interpolation of this PSF increases the sampling to a hundredth of a pixel, which is the accuracy to which we estimate the position from the $\chi^{2}$ grid. At this point, we perform cosmic ray filtering by removing images from the sample with a larger than 10$\sigma$ outlier in the residuals between image and best-fit PSF model. Because of the high cadence of the EPOCh observations, we simply reject any images containing a cosmic ray overlying the stellar PSF from the time series.

Flat fields for the High Resolution Imager were last taken while the spacecraft was on the ground. We construct secondary flat fields using a ``stimulator'' LED on board. Because this LED is mounted to one side of the CCD, we first remove an overall trend of increasing brightness with position by dividing out a best-fit plane to the flat field. We then produce a final ``stim'' by coadding the 100 stims taken in each block of stim observations, using pixel-by-pixel outlier rejection to account for cosmic ray hits. We observed changes to the stim in the form of dark spots continuously during observations; we attribute this to possible radiation damage. However, the standard deviation of the time series was unchanged when we removed observations for which one of these dark spots was present in the aperture.

After we use PSF fitting to determine the position of the star on the CCD, we use aperture photometry to perform the photometric extraction. The structure of the CCD includes two rows in the middle of the array with physically smaller pixels, such that the PSF has a different shape when the star straddles these rows. We account for this problem by scaling these rows by the multiplicative factor that best recovers a flat dependence of out-of-transit flux on row position. We also perform a similar multiplication by a scale factor for the two middle columns. In the first version of the photometry pipeline, we also divided out a spline fit to the flux, first as a function of y position, and then as a function of x, to account for additional flux dependence on position. We decided on an optimal aperture radius based on analysis of the standard deviation of the out-of-transit time series. We found that this standard deviation was minimized for an aperture radius of 10 pixels, corresponding to twice the HWHM of the PSF. 

After performing the stim division and the scaling of the two middle columns and rows, the time series still suffered from significant red noise. At this point, we implemented a 2D spline fit to the data with position, which largely removed the red noise when divided from the time series. Figure 1 shows the time series before and after this 2D spline correction. In the bottom panel, we show that the time series after the 2D spline bins down roughly as predicted for Gaussian noise. In the corrected time series, we attribute the scatter to photon noise and low-level cosmic rays.

\begin{figure}
% \vspace*{-2.0 cm}
\begin{center}
 \includegraphics[width=5in]{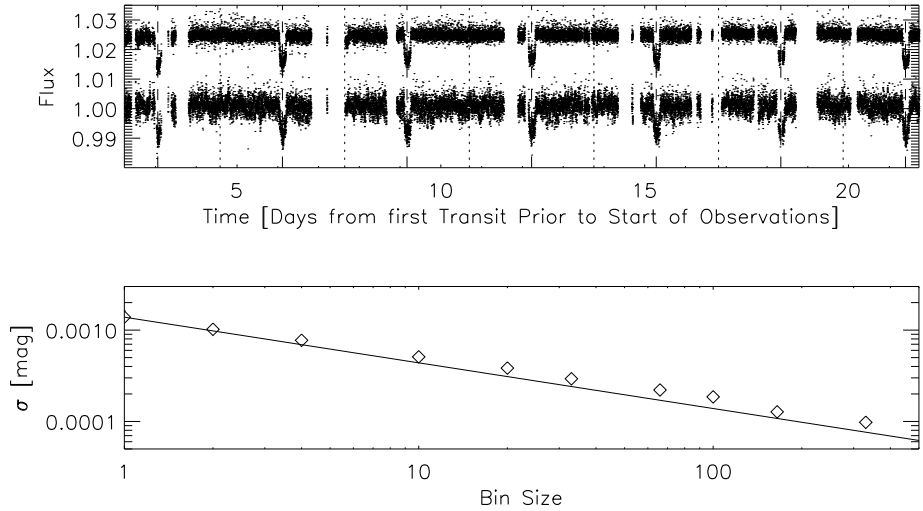} 
% \vspace*{-1.0 cm}
 \caption{\textit{Top panel:} HAT-P-4 time series before (lower curve) and after (upper curve) 2D spline correction. The uncorrected time series has had the two middle rows and columns in each image scaled by a multiplicative factor to reduce the flux dependence on position, but the images were not divided by a flat field constructed from a stim. Instead, we have used the 2D spline to correct for interpixel variations. \textit{Bottom panel:} The data (diamond symbols) bin down consistently with the expectation for Gaussian noise (shown with a line, renormalized to match the value at N=1).}
   \label{fig1}
\end{center}
\end{figure}

\begin{figure}
% \vspace*{-2.0 cm}
\begin{center}
 \includegraphics[width=5.2in]{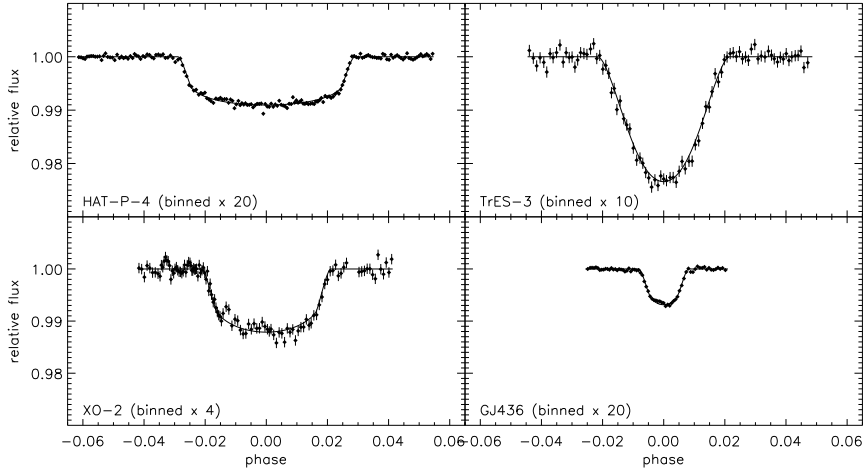} 
% \vspace*{-1.0 cm}
 \caption{Phased and binned light curves for the four EPOCh targets observed to date. We have applied the 2D position decorrelation to HAT-P-4 and TrES-3. The solid curve corresponds to a best-fit transit curve solution to these data, using \cite[Mandel \& Agol (2002)]{Mandel02} routines for generating light curves. The limb darkening coefficients were assumed for HAT-P-4 and TrES-3 from stellar models, but were allowed to vary for XO-2 and GJ 436.}
   \label{fig2}
\end{center}
\end{figure}

\section{Preliminary Results and Discussion}
We have observed seven transits of HAT-P-4, five transits of TrES-3, three partial transits of XO-2, and seven transits of GJ 436. Figure 2 shows the EPOCh observations, phase-folded and binned, for these four systems. Our strategy is to conduct the science analysis of the data while working in parallel to refine the photometric precision through improvements to our extraction pipeline. We are currently conducting a Markov Chain Monte Carlo modeling and reflected light analysis to refine the system parameters and constrain the planetary albedos. These analyses are explained in more detail in Christiansen et~al. (these proceedings). We are also conducting preliminary searches for shallow transits made by additional bodies. For GJ 436, this search is especially interesting, since the eccentricity of the known transiting planet points to the possible presence of a second planetary companion. We are just beginning to develop the software to search for these additional transits, which will take steps briefly outlined here. 

The set of lightcurves for additional transiting bodies is a three parameter family if we make the simplifying assumption of no limb darkening of the host star and an inclination angle $i$ of 90$^{\circ}$: radius of the planet $R_{p}$, period of the planet $P$, and phase $\phi$. Using the Mandel and Agol (2002) routines for generating lightcurves given these parameters, we compute a grid of models corresponding to additional possible planets. For each of these models, we compute the $\chi^{2}$, using the out-of-transit standard deviation to estimate the error in each point. At the positions of local minima, we will change to a routine that more finely samples the $\chi^{2}$ space so that we can accurately gauge the significance levels of the parameters. The improvement in the $\chi^{2}$ over the null hypothesis will also determine whether we can claim a positive detection.  The photometric precision of the GJ 436 time series (with S/N=44 for each transit event) would enable a detection of a 1.3 R$_{\oplus}$ planet with S/N of 4 and a 2.1 R$_{\oplus}$ with S/N of 10, even if the planet produced only a single transit event.

\end{document}